\documentclass{PoS}
\usepackage{subfigure}

\title{$\mathbf{K^0-\bar{K}^0}$ Mixing Beyond the SM from $\mathbf{N_f=2}$ tmQCD}

\ShortTitle{$K^0-\bar{K}^0$ Mixing Beyond the SM from $N_f=2$ tmQCD}
\author{ETM Collaboration}
\author{\speaker{P.~Dimopoulos}, G.~Martinelli \\
        Dip. di Fisica, Universit\`a di Roma ``La Sapienza'' \\
        Piazzale Aldo Moro 2, I-00185 Rome Italy \\ 
        E-mail: \email{dimopoulos@roma2.infn.it, martinelli@roma1.infn.it}}
\author{R.~Frezzotti, G.C.~Rossi, A.~Vladikas$^{\dagger}$\\
        Dip. di Fisica, Universit\`a di Roma ``Tor Vergata",\\
        $^{\dagger}$ INFN-``Tor Vergata"\\
        Via della Ricerca Scientifica 1, I-00133 Rome, Italy \\
        E-mail: \email{\{frezzotti,rossig,vladikas\}@roma2.infn.it}}

\author{V.~Gimenez\\
        Dep. de Fisica Te\`orica and IFIC, Univ. de Val\`encia-CSIC,\\
        Dr.~Moliner 50, E-46100 Val\`encia, Spain\\
        E-mail: \email{vicente.gimenez@uv.es}}

\author{V.~Lubicz, S.~Simula$^{\ddagger}$\\
         Dip. di Fisica, Universit{\`a} di Roma Tre,\\ 
         $^{\ddagger}$ INFN-Roma Tre \\
         Via della Vasca Navale 84, I-00146 Rome, Italy \\
         E-mail: \email{lubicz@fis.uniroma3.it, simula@roma3.infn.it}}

\author{F.~Mescia\\
        Departament d'Estructura i Constituents de la Mat\`eria and
        Institut de Ci\`encies del Cosmos (ICCUB), Universitat de Barcelona, Diagonal 647, 
        08028 Barcelona, Spain \\      
        E-mail: \email{mescia@ub.edu}}
\author{M.~Papinutto\\
        Laboratoire de Physique Subatomique et de Cosmologie, UJF/CNRS-IN2P3/INPG,\\
        53 rue des Martyrs, 38026 Grenoble, France \\
        E-mail: \email{Mauro.Papinutto@lpsc.in2p3.fr}}

\abstract{We present preliminary results on the of neutral kaon oscillations in
extensions of the Standard Model. Using  $N_f=2$ maximally twisted sea
quarks and Osterwalder-Seiler valence quarks, we achieve both
O(a)-improvement and continuum-like renormalization pattern for the relevant
four-fermion operators. We perform simulations at three values of the
lattice spacing and extrapolate/interpolate our results to the continuum
limit and physical light/strange quark mass. The calculation of the
renormalization constants of the complete  operator basis is performed non-
perturbatively in the RI-MOM scheme.
}

\FullConference{The XXVIII International Symposium on Lattice Field Theory--LATTICE2010\\
		 June 14-19  2010 Villasimius, Sardinia, Italy }

\begin{document}

\section{Introductory Remarks and Calculation Setup}

Flavour Changing Neutral Currents (FCNC) and CP violation may furnish
useful information  on the impact of models defined beyond the standard model
(BSM). 
In various BSM models (like for example the supersymmetric ones) there appears the possibility
for $\Delta S = 2$ processes at one loop, even mediated by the strong interactions. These
effects are thus potentially large. The computation of the relevant matrix elements of the
effective Hamiltonian in combination with the experimental value of $\varepsilon_K$ offers the
chance to constrain the values of the model parameters (like for instance the off-diagonal terms
of the squark mass matrix in supersymmetric models \cite{Ciuchini-etal98}) which enter explicitly in the Wilson
coefficients.

In all the BSM models the effective Hamiltonian relevant for the $\Delta S=2$ processes takes the general form
\begin{equation}
{\cal H}_{\rm{eff}}^{\Delta S=2} = \sum_{i=1}^{5} C_i(\mu) {\cal O}_i 
+ \sum_{i=1}^{3} \tilde{C}_i(\mu) \tilde{{\cal O}}_i 
\end{equation}
where the operators ${\cal O}_i$ are defined by
\begin{eqnarray}
{\cal O}_1 &=& [\bar{s}^a \gamma_\mu (1-\gamma_5)d^a][\bar{s}^b \gamma_\mu (1-\gamma_5)d^b], \nonumber \\
{\cal O}_2 &=& [\bar{s}^a (1-\gamma_5)d^a][\bar{s}^b  (1-\gamma_5)d^b], ~~~~~~~~
{\cal O}_3 = [\bar{s}^a (1-\gamma_5)d^b][\bar{s}^b  (1-\gamma_5)d^a], \nonumber \\
{\cal O}_4 &=& [\bar{s}^a (1-\gamma_5)d^a][\bar{s}^b  (1+\gamma_5)d^b], ~~~~~~~~
{\cal O}_5 = [\bar{s}^a (1-\gamma_5)d^b][\bar{s}^b  (1+\gamma_5)d^a] 
\end{eqnarray}
We note that in the SM case only the operator ${\cal O}_1$ contributes.
The parity-even parts of the operators 
\begin{eqnarray}
\tilde{{\cal O}}_1 &=& [\bar{s}^a \gamma_\mu (1+\gamma_5)d^a][\bar{s}^b \gamma_\mu (1+\gamma_5)d^b], \nonumber \\
\tilde{{\cal O}}_2 &=& [\bar{s}^a (1+\gamma_5)d^a][\bar{s}^b  (1+\gamma_5)d^b], ~~~~~~~~
\tilde{{\cal O}}_3 = [\bar{s}^a (1+\gamma_5)d^b][\bar{s}^b  (1+\gamma_5)d^a]  
\end{eqnarray}
coincide with those of the operators ${\cal O}_i$. Therefore, due the parity conservation 
of the strong interactions only the parity-even contributions of the operators ${\cal O}_i$ need to be calculated.  
Defining a basis of the parity even operators as follows
\begin{eqnarray} \label{LAT_op}
O^{VV} &=& (\bar{s}\gamma_{\mu}d)(\bar{s}\gamma_{\mu}d), ~~~~~~~~ 
O^{AA} = (\bar{s}\gamma_{\mu}\gamma_5 d)(\bar{s}\gamma_{\mu}\gamma_5 d), \nonumber \\
O^{PP} &=& (\bar{s}\gamma_5 d)(\bar{s}\gamma_5 d), ~~~~~~~~~\, 
O^{SS} = (\bar{s}d)(\bar{s}d), \nonumber \\
O^{TT} &=& (\bar{s}\sigma_{\mu \nu} d)(\bar{s} \sigma_{\mu \nu} d) 
\end{eqnarray}
through a Fierz transformation we obtain
\begin{eqnarray} \label{BSM_op}
{\cal O}_1 &=& (O^{VV} + O^{AA}), ~~~~~~ 
{\cal O}_2 = (O^{SS} + O^{PP}),   ~~~~~~ 
{\cal O}_3 = -\frac{1}{2}(O^{SS} + O^{PP} - O^{TT}),  \nonumber \\
{\cal O}_4 &=& (O^{SS} - O^{PP}), ~~~~~~~
{\cal O}_5 = -\frac{1}{2} (O^{VV} - O^{AA})   
\end{eqnarray}

Up to now,  lattice calculations have been presented in the quenched approximation 
(\cite{Donini:1999nn}, \cite{Babich_etal06}, \cite{Nakamura_etal06}) with the
exception of a preliminary study of the bare matrix elements using unquenched simulations 
with 2+1 dynamical quarks \cite{Wennekers:2008sg}.
 
Our lattice computations have been performed at three values of the lattice spacing  
using the $N_f=2$ dynamical quark configurations produced by the ETM collaboration~\cite{Baron:2009wt}.
ETMC dynamical configurations have been produced with the tree-level Symmanzik
improved action in the gauge sector while the dynamical quarks have been regularized 
by employing the twisted mass (tm) formalism \cite{tmQCD1}. It has been demonstrated that 
with the condition at {\it maximal twist} this formalism provides automatic 
$O(a)$-improved physical quantities \cite{Frezz-Rossi1}. 

In the so called {\em physical} basis the fermion lattice action concerning the sea sector is written

\begin{equation}\label{sea-action}
S_{sea}^{\rm Mtm} =  a^4 \sum_x \bar \psi (x) \Big \{ \frac{1}{2} \sum_\mu \gamma_\mu(\nabla_\mu
+ \nabla^\ast_\mu ) - i \gamma_5 \tau^3 \big [ M_{\rm cr} -
\frac{a}{2}
\sum_\mu \nabla^\ast_\mu \nabla_\mu \big ] + \mu_{sea} \Big \} \psi(x)
\end{equation}
where the Wilson's $r$ parameter has been set to unity,
$\psi(x)$ is the quark flavour doublet, $\nabla_\mu$ and $\nabla_\mu^\ast$ are
nearest-neighbour forward
and backward lattice covariant derivatives, $\mu_{sea}$ is the (twisted) sea quark mass and $M_{\rm cr}$ the
critical mass.
It has been shown that the use of the tm regularization can simplify the renormalization pattern 
properties of the four-fermion operators (e.g. $B_{\rm{K}}$) \cite{tmQCD1, AlphaBK, PenSinVla}. Moreover,
both $O(a)$ improvement and  continuum-like operator renormalization pattern can be achieved 
introducing a valence quark action of the Osterwalder-Seiler type \cite{Osterwalder:1977pc} by allowing 
for a replica of the down ($d$, $d'$) and strange ($s$, $s'$)
flavours~\cite{Frezz-Rossi2}. The valence quark action   assumes the form 
\begin{equation} \label{action}
S_{val} = a^4 \sum_{x} \sum_{f=d,d',s,s'} \bar{q}_f(x) \,
\Big \{ \frac{1}{2} \sum_\mu \gamma_\mu(\nabla_\mu
+ \nabla^\ast_\mu ) - i \gamma_5 r_f \big [ M_{\rm cr} -
\frac{a}{2}
\sum_\mu \nabla^\ast_\mu \nabla_\mu \big ] + \mu_{f} \Big \}
\, q_f(x)
\end{equation}
with $ -r_s = r_d = r_{d'} =r_{s'} = 1$. Note that the field $q_f$ represents just one individual flavour. The  
four fermion operators of Eq.~(\ref{LAT_op}) can be written in general form  as
$O_{\Gamma \tilde{\Gamma}} =  2 \{ [\bar{q_1} \Gamma q_2][\bar q_3 \tilde{\Gamma} q_4]+ (q_2 \leftrightarrow q_4)\}$ with $q_1$ and $q_3$  identified with the strange quark (by setting $\mu_s=\mu_{s'}=\mu_{strange}$) and
$q_2$ and $q_4$   identified with the down quark (by setting $\mu_d=\mu_{d'}=\mu_{\ell}$); 
the interpolating fields for the 
external (anti)Kaon states are made up of a tm-quark pair ($\bar{d}\gamma_5 s$,
with $-r_s = r_d$) and a OS-quark pair ($\bar{d}'\gamma_5 s'$, with $r_{d'} =r_{s'}$).
This {\it mixed} action setup with maximally twisted Wilson-like quarks has been studied in
detail in Ref.~\cite{Frezz-Rossi2} and it has been demonstrated that it  allows for 
an easy matching of sea and valence quark masses and leads to unitarity violations that vanish as $a^2$
as the continuum limit is approached. Moreover in the present computation the quark mass matching is incomplete
because we are neglecting the sea strange quark (i.e. we work in a partially quenched set-up). 
A first test that the proposed method leads to  $O(a)$ improved results 
was already performed in the calculation of $B_K$ with fully quenched quarks \cite{ALPHA-BK-2009}.
In a recent publication \cite{Constantinou:2010qv} our collaboration, using the OS-tm mixed action set-up, has presented an 
$O(a)$-improved computation of $B_{\rm{K}}$ with $N_f=2$ dynamical quarks. 
Using non-perturbative operator renormalisation and three values for the lattice spacing, the RGI value of $B_{\rm{K}}$  in the
continuum limit is $ B_{\rm{K}}^{\rm{RGI}} = 0.729 \pm 0.030$.  
  
\begin{table}[!h]
\begin{center}
\begin{tabular}{cccccccccc}
\hline \hline
 $\beta$  &&  $a^{-4}(L^3 \times T)$ && $a\mu_{\ell}~=~a\mu_{sea}$      &&  $a\mu_{\rm{``}s\rm{"}}$ &&   \\
\hline
3.80      &&  $24^3 \times 48$&& 0.0080 0.0110   && 0.0165, 0.0200, 0.0250   &&     \\
($a\sim0.1~\mbox{fm}$)          &&                  &&                 &&    &&     \\
\hline
3.90      &&  $24^3 \times 48$&& 0.0040, 0.0064 &&   0.0150, 0.0220, 0.0270 &&    \\
          &&                  && 0.0085, 0.0100 &&           &&    \\
''        &&  $32^3 \times 64$&& 0.0030, 0.0040 &&   0.0150, 0.0220, 0.0270 &&    \\
($a\sim0.085~\mbox{fm}$) && && && && \\
\hline
4.05      &&  $32^3 \times 64$&& 0.0030, 0.0060 &&   0.0120, 0.0150, 0.0180 &&      \\
($a\sim0.065~\mbox{fm}$)          &&                  && \hspace*{-1.3cm} 0.0080         &&    &&      \\
\hline \hline
\end{tabular}
\end{center}
\caption{Simulation details}
\label{simuldetails}
\end{table}

In Table~\ref{simuldetails} we give the simulation details and the values of the sea and the valence quark
masses at each value of the gauge coupling for the calculation presented in this work.
The smallest sea quark mass corresponds to a pion of about 280 MeV 
for  the case of $\beta=3.90$. For $\beta=4.05$ the lightest pion weighs 300 MeV while
for $\beta=3.80$ the lowest pion mass is around 400 MeV.
The largest sea quark mass for the three values of the lattice spacing is about
half the strange quark mass.
For the inversions in the valence sector
we have made use of the stochastic method (one--end trick of Ref.~\cite{Michael}) in order
to increase the statistical information. Propagator sources have been located  at randomly 
chosen timeslices. For more details on the dynamical configurations and the stochastic 
method application see Ref.~\cite{etmc-light}.

\section {B-Parameters and Four-Fermion Matrix Elements}
As it has been shown in \cite{Frezz-Rossi2}, the discrete symmetries  guarantee 
that in the OS-tm mixed action
set-up the renormalisation of the four-fermion operators is continuum-like in the sense that the mixing
between operators of different naive chirality 
is of order $O(a^2)$ or higher.
An equivalent view of the same property can be offered by the fact that in the (unphysical) tm-basis the parity-even part of each of the four fermion operators is mapped over its parity-odd counterpart. Then due to the CPS 
symmetries  the parity odd operators have the same block-diagonal renormalisation matrix pattern 
both in the continuum and at finite value of the lattice spacing (\cite{Bernard}, \cite{rimom4}).    

\noindent The B-parameters for the operators~(\ref{BSM_op}) are defined as
\vspace*{-0.1cm} 
\begin{eqnarray}
 \langle \bar{K}^{0} | {\cal O}_1(\mu) | K^{0} \rangle &=& B_1(\mu) \frac{8}{3} m_{K}^{2} f_{K}^{2}  \equiv B_K(\mu)
  \frac{8}{3} m_K^2 f_K^2 \nonumber \\
  \langle \bar{K}^{0} | {\cal O}_i(\mu) | K^{0} \rangle &=&  C_i B_i(\mu) 
\LSB \frac{ m_{K}^{2} f_{K}}{ m_s (\mu) + m_d(\mu)} \RSB^2, \nonumber 
\end{eqnarray}
where $C_i = \{-5/3, 1/3, 2, 2/3\}, ~~ i=2, \ldots, 5$.
The matrix element of the operator ${\cal O}_1$ vanishes in the chiral limit while the  matrix element of the 
operators ${\cal O}_i~ i=2,\ldots,5$ get a non-zero value in the chiral limit. From the above equations 
it can be seen that the calculation of the $B_i$ parameters for $i=2,\ldots,5$ involves the calculation of the 
quark mass at the same scale $\mu$.  In order to avoid any extra systematic uncertainties 
in the computation of  the matrix elements due to the  quark mass
evaluation,  it has been proposed the calculation of appropriate ratios of the four-fermion matrix elements
(\cite{Donini:1999nn}, \cite{Babich_etal06}). Here, besides the calculation of the $B$ parameters, we also consider the following ratios
\vspace*{-0.3cm}
\begin{equation} 
R_i =  \Big{(}\frac{f_{K}^{2}}{m_{K}^{2}}\Big{)}_{\rm{exp}} \LSB   \Big{(}\frac{m_K}{f_K}\Big{)}_{-r_s = r_d} 
\Big{(} \frac{m_K}{f_K}\Big{)}_{r_{d'} =r_{s'}}
\frac{\ME{\bar{K}^0}{{\cal O}_i(\mu)}{K^0}}{\ME{\bar{K}^0}{{\cal O}_1(\mu)}{K^0}}    \RSB ~~~ i=2,\ldots, 5
\label{RATIO_O}
\end{equation} 

\vspace*{-0.2cm}
\begin{figure}[!h]
\begin{center} \hspace*{-0.8cm}
\includegraphics[scale=0.57, angle=-90]{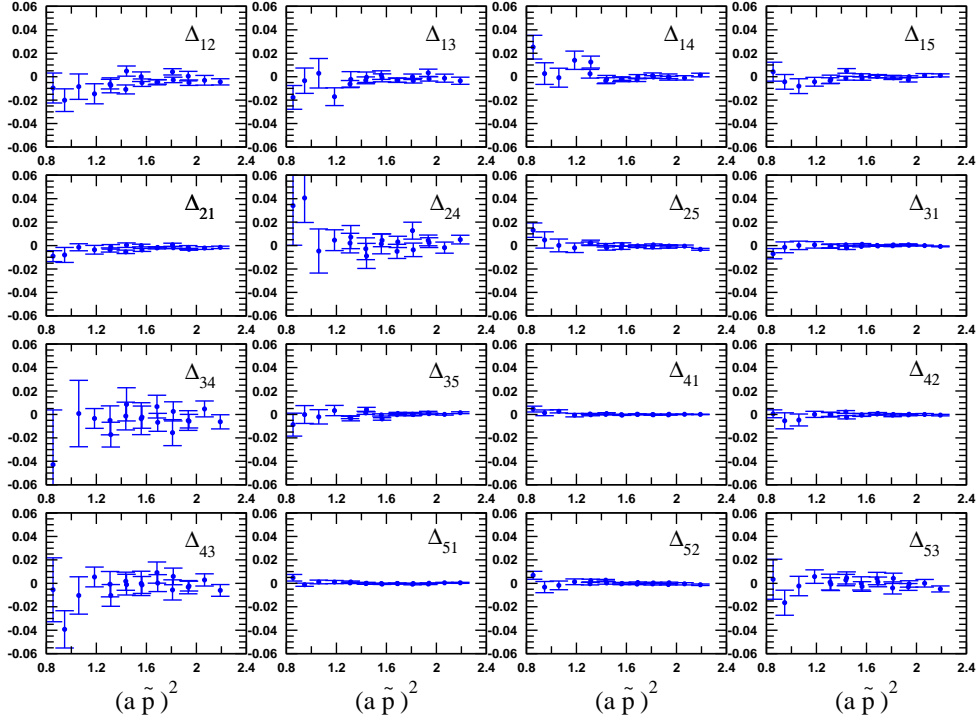}
\vspace*{-0.3cm}\caption[]{The off-diagonal off-block elements of the four-fermion RC-matrix operator, $\Delta{ij}$ , (for
$\beta = 3.90$) which take values compatible with zero. }
\label{fig:Mixing}
\end{center}
\end{figure}

\begin{figure}[!h]
\begin{center}
\subfigure[]{\includegraphics[scale=0.50]{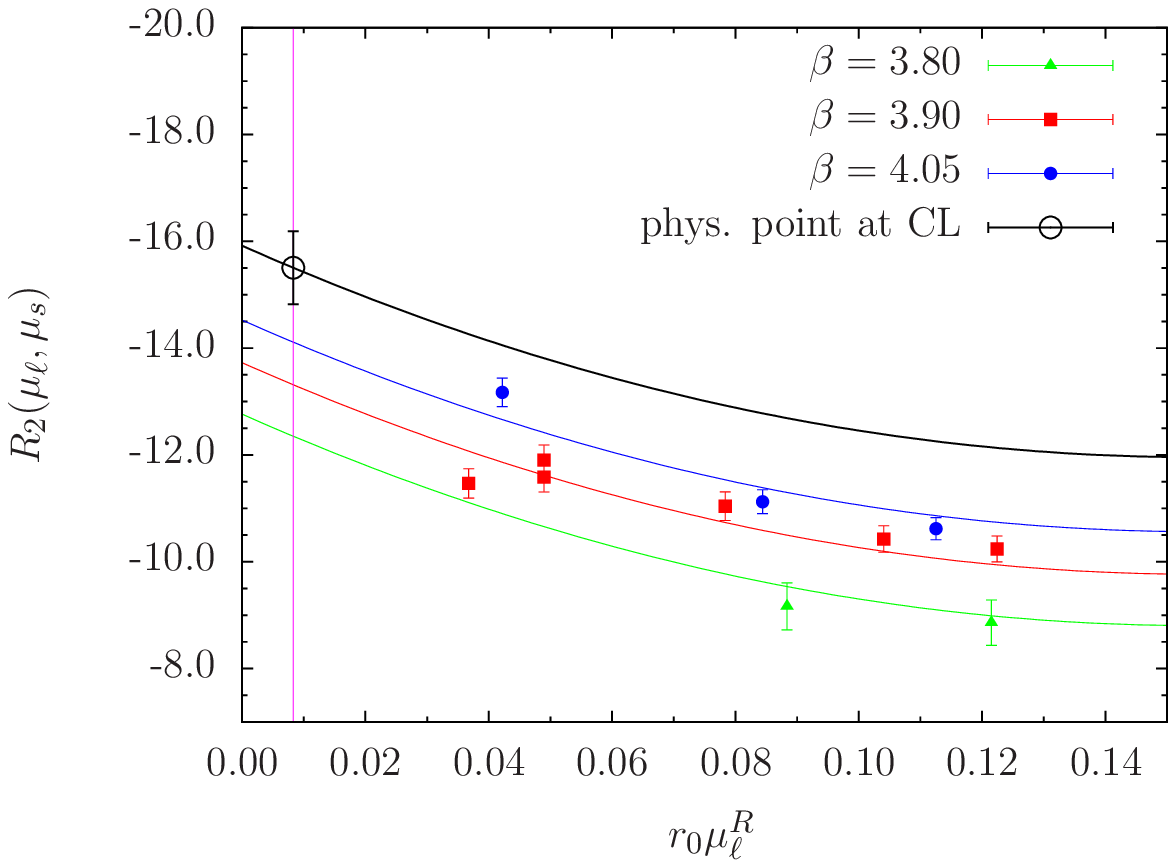}}
\subfigure[]{\includegraphics[scale=0.50]{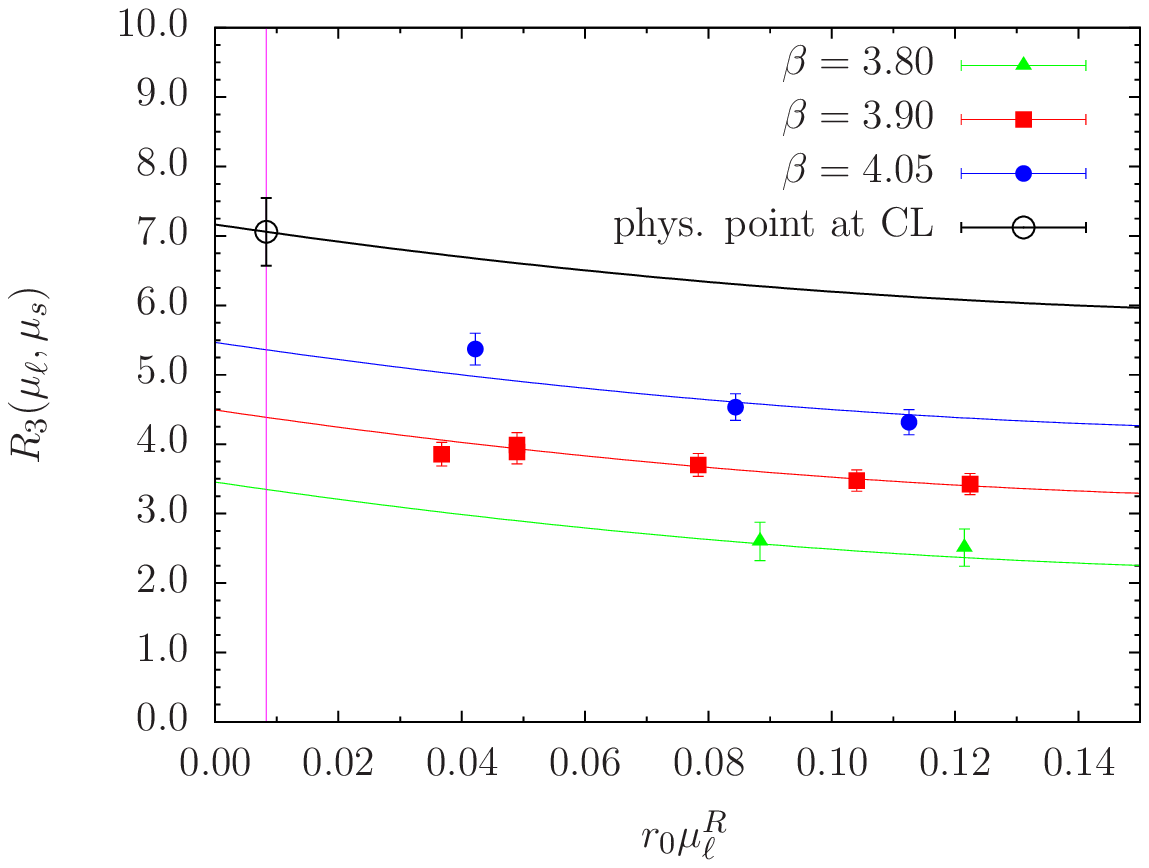}}
\subfigure[]{\includegraphics[scale=0.50]{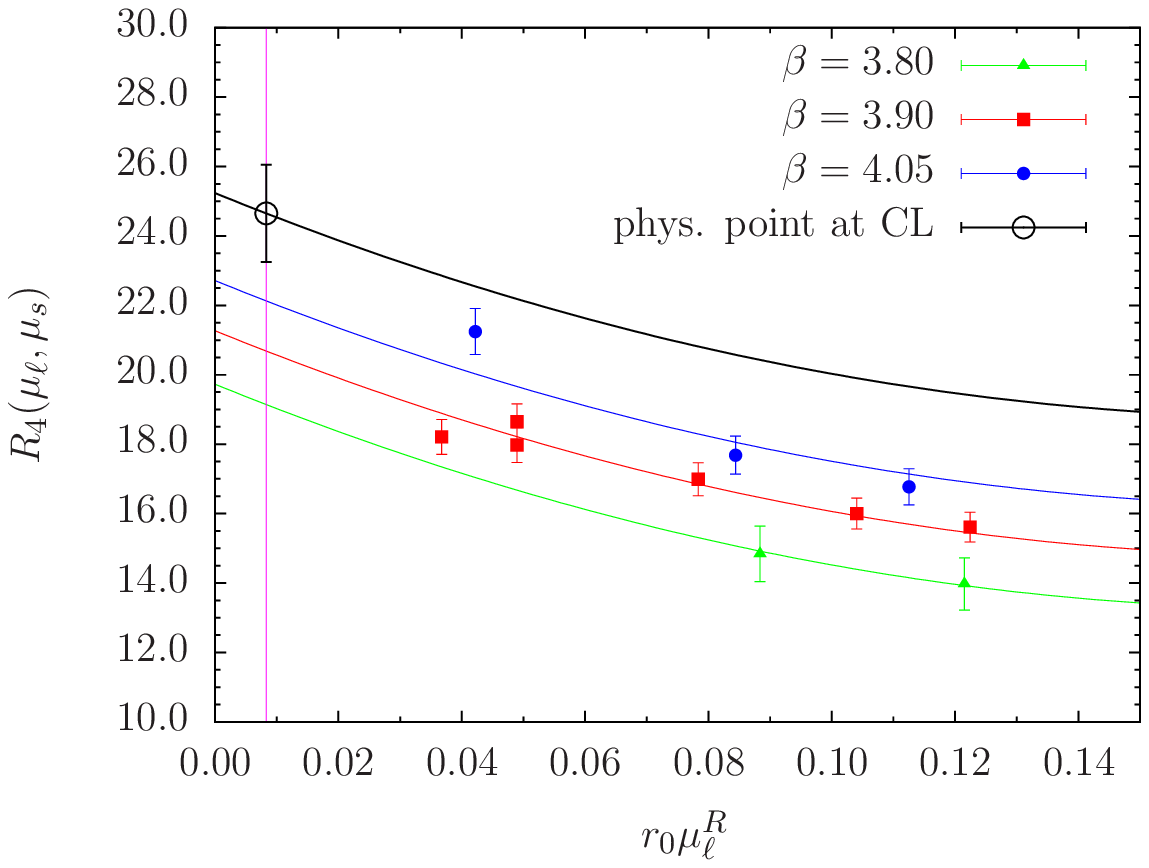}}
\subfigure[]{\includegraphics[scale=0.50]{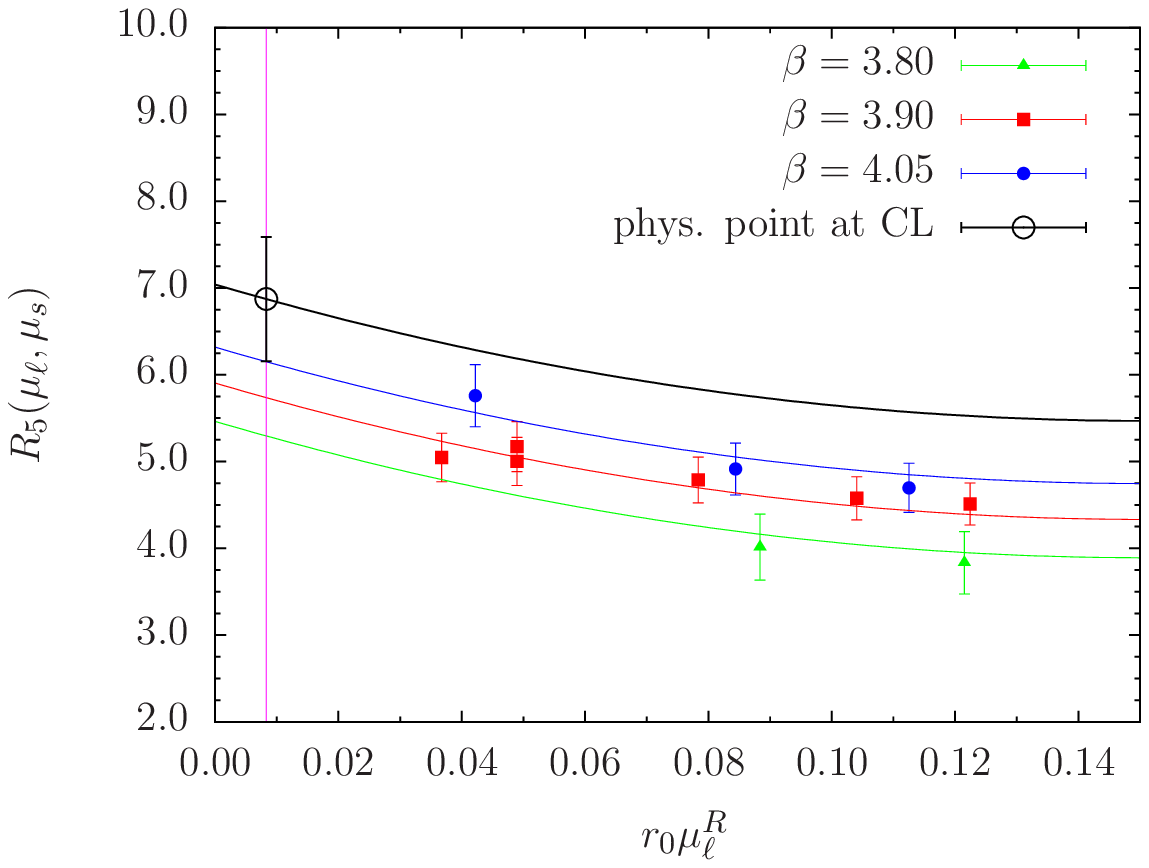}}
\vspace*{-0.5cm}\caption[]{Combined fits of $R_i(\mu_{\ell},\mu_{s}), ~ i=2, \ldots, 5$ 
with respect to the renormalised light quark mass, 
$r_0 \mu_{\ell}^{R}$,  in the $\overline{MS}$ scheme. The fit functions, shown here, are second order polynomial 
functions of the renormalised light quark mass with the addition of a term proportional to $a^2$.  
}
\label{fig:RATIOS}
\end{center}
\end{figure}

The computation of the renormalisation constants (RCs) relevant for both the four-fermion 
and two-fermion operators\footnote{In our 
OS-tm mixed action set-up we need to use the RCs for the scalar and the pseudoscalar density operators in the 
calculation of $B_i$ for $i=2,\ldots,5$. For  $B_1$, instead,  (i.e. $B_{\rm{K}}$) the normalisation constants for 
the axial and the vector current are needed.} has been performed in a non-perturbative way  
using the RI-MOM scheme following the strategies detailed in Refs.~{\cite{rimom2} and \cite{Constantinou:2010gr}. In Fig.~\ref{fig:Mixing} we show, for $\beta=3.90$, 
that all the off-diagonal off-block elements of the four-fermion RC-matrix operator, $\Delta_{ij}$,
which are expected to vanish with the tm-OS mixed action setup, take values compatible with
zero.
In Fig.~\ref{fig:RATIOS} we show the combined fits of $R_i(\mu_{\ell},\mu_{s}), ~ i=2, \ldots, 5$ 
with respect to the renormalised light quark mass, $r_0 \mu_{\ell}^{R}$,  in the $\overline{MS}$ scheme.

\begin{table}[!h]
\begin{center}
\begin{tabular}{ccccc}
\hline \hline
fit function  & $i$      & $\frac{<{\cal O}_{i}>}{<{\cal O}_{1}>}$ (using $R_i$)  
& $\frac{<{\cal O}_{i}>}{<{\cal O}_{1}>}$ (using $B_i$) & $B_i$          \\
\hline \hline
&&&& \\
quadratic     &   2              & -15.5(1.0)         &  -17.1(1.5)&  0.56(0.04)  \\
              &   3              &  ~~~7.1(0.5)       &   ~~~8.7(0.9) &  1.43(0.13)  \\
              &   4              &  24.6(1.4)         &   27.9(2.5)&  0.76(0.06)   \\
              &   5              &  ~~~6.9(0.7)       &   ~~~7.6(1.2) &  0.63(0.09)  \\
&&&& \\
\hline
&&&& \\
linear        &   2              & -15.0(0.6)         & -17.1(1.2) &  0.56(0.02) \\
              &   3              &  ~~~7.0(0.3)       &   ~~~8.8(0.6) &  1.44(0.08)   \\
              &   4              &  24.2(0.9)         &  27.8(1.8) &  0.76(0.04)  \\
              &   5              &  ~~~6.6(0.5)       &  ~~~7.5(0.7)  &  0.62(0.06)  \\
\hline \hline
\end{tabular} \vspace*{0.3cm}
\vspace*{-0.3cm} \caption{Preliminary results in the continuum limit  for the B-parameters and the
ratios $\frac{<{\cal O}_{i}>}{<{\cal O}_{1}>},~ i=2, \ldots, 5$.
calculated at the physical point ($\mu_d$, $\mu_s$, $a=0$). All results are given in the $\overline{MS}$ scheme.
The ratios of the operators' matrix elements  are estimated
using either the direct (3rd column)  or and  the indirect method i.e. through the $B_i$ calculation (4th column) . 
}
\label{OisuO1}
\end{center}
\end{table}

In Table~\ref{OisuO1} we present our preliminary results in the continuum limit and in the $\overline{MS}$ 
scheme for the B-parameters and the
ratios $\frac{<{\cal O}_{i}>}{<{\cal O}_{1}>},~ i=2, \ldots, 5$ 
calculated at the physical point $(\mu_{d}, \mu_{s})$. The ratios,  
$\frac{<{\cal O}_{i}>}{<{\cal O}_{1}>},~ i=2, \ldots, 5$, have been calculated either directly 
(through $R_i$) or using the $B_i$ estimates and the values of the u/d and strange quark mass~\cite{qmasses}.  
The results are  compatible within one or two standard 
deviations. We have tried fit functions using either a second  or first order polynomial with respect 
to the light quark mass to which a term proportional to $a^2$ has been added; we do not notice a significant difference in the final continuum limit values. 
We should note that the use of a fit function containing a NLO logarithmic term 
leads to rather similar results with those obtained with a second order polynomial fit function.

\section*{Acknowledgements}
 V. G. thanks the MICINN (Spain) for partial support under grant FPA2008-03373 and the
Generalitat Valenciana (Spain) for partial support under grant GVPROMETEO2009-128.
M.P. acknowledges financial support by a Marie Curie European
Reintegration Grant of the 7th European Community Framework
Programme under contract number PERG05-GA-2009-249309.


\end{document}